\documentclass[preprint]{aastex631}

\shorttitle{Force-freeness of solar photosphere}
\shortauthors{M. Zhang \& H.C. Zhang}

\graphicspath{{./}{figures/}}

\begin{document}

\title{The force-freeness of the solar photosphere: Revisit with new approach and large datasets}

\author[0000-0002-3141-747X]{Mei Zhang}
\affiliation{National Astronomical Observatories, Chinese Academy of Sciences, Beijing 100101, China; zhangmei@nao.cas.cn}
\affiliation{School of Astronomy and Space Sciences, University of Chinese Academy of Sciences, Beijing 100049, China}
\affiliation{High Altitude Observatory, National Center for Atmospheric Research, 3080 Center Green Drive, Boulder, CO 80301, USA}

\author{Haocheng Zhang}
\affiliation{Jericho High School, 99 Cedar Swamp Road, Jericho, NY 11753, USA}

\begin{abstract}
Although it is generally believed that the solar photosphere is not magnetically force-free owning to its high plasma $\beta$, the estimations of force-freeness using observed magnetograms have produced disputable results. Some studies confirmed that the photosphere is largely not force-free whereas some authors argued that the photosphere is not far away from being force-free. In a previous paper of ours we demonstrated that, due to the fact that the noise levels of the transverse field in the magnetograms are much larger than those of the vertical field, wrong judgements on the force-freeness could be made: a truly force-free field could be judged as being not-force-free and a truly not-force-free field could be judged as being force-free.  Here in this letter we propose an approach to overcome this serious problem. By reducing the spatial resolution to lower the noise level, the heavy influence of the measurement noise on the force-freeness judgement can be significantly suppressed. We first use two analytical solutions to show the success and the effectiveness of this approach. Then we apply this new approach to two large datasets of active region magnetograms, obtained with the HMI/SDO and SP/Hinode, respectively. Our analysis shows that the photospheric magnetic fields are actually far away from being force-free. Particularly and most notably, the mean value of $F_z/F_p$ (where $F_z$ the net Lorentz force in the vertical direction and $F_p$ the total Lorentz force) is as low as $-0.47$, with more than $98\%$ of the active regions having $|F_z/F_p|>0.1$, when using the SP/Hinode magnetograms of true field strength.
\end{abstract}

\keywords{The Sun --- Solar Photosphere --- Solar Magnetic Fields}

\section{Introduction} 

It is well known that the Sun’s magnetic field is the engine and energy source driving all solar phenomena \citep{Charbonneau2014}. These phenomena collectively define the solar activity, including those violent eruptions such as coronal mass ejections \citep{Zhang2005}. An accurate understanding on how the solar magnetic fields are produced and evolved is then very crucial. However, so far only the photospheric magnetic field can be measured with fair accuracy and reasonable spatial and temporal resolutions. The large bodily coronal magnetic fields are mainly estimated from nonlinear force-free field extrapolations \citep{Wiegelmann2021}. 

It is a common practice that the observed photospheric magnetograms are used as the boundary conditions in the force-free field extrapolations. Then as a first step, whether the observed photospheric magnetograms satisfy the force-free condition or not needs to be checked.  \citet{Gary2001} pointed out that the solar photosphere has its $\beta>1$, so it is likely that the photosphere is not force-free.

\citet{Low1985} proposed a method to check whether the observed magnetograms satisfy the force-free condition or not. He pointed out that, in an isolated magnetic structure, a necessary condition for a force-free field to exist above a measured layer is:
\begin{equation}
F_x \ll F_p, \,\, F_y \ll F_p, \,\, F_z \ll F_p  ~,
\end{equation}
\noindent where $F_x$, $F_y$ and $F_z$ are the components of the net Lorentz force, and $F_p$  is the characteristic magnitude of the total Lorentz force that can be brought to bear on the atmosphere if the magnetic field is not force-free. These components of net Lorentz force can be estimated, using \citet{Low1985} method from the observed vector magnetograms, by following formula:
\begin{equation}
                   F_{x} = - \frac 1{4\pi} \int B_{x} B_{z} dxdy~,
                 \end{equation}
                 \begin{equation}
                    F_{y} = - \frac 1{4\pi} \int B_{y} B_{z} dxdy~,
                 \end{equation}
                 \begin{equation}
                    F_{z} = - \frac 1{8\pi} \int (B_{z}^2 - B_{x}^2- B_{y}^2) dxdy~,
                 \end{equation}
                 \begin{equation}
                    F_{p} =  \frac 1{8\pi} \int (B_{z}^2 + B_{x}^2+ B_{y}^2) dxdy~,
                 \end{equation}
 \noindent where $B_x$, $B_y$ and $B_z$ are the three components of the vector magnetic field, in which $B_z$ is the vertical magnetic field and
$B_x$, $B_y$ are the two components of the horizontal magnetic field $B_t$.

\citet{Metcalf1995} presents the first study of using \citet{Low1985} method to estimate the force-freeness of an active region NOAA 7216. It is also the first study where a value of $|F_x/F_p|$, $|F_y/F_p|$ and $|F_z/F_p|$ being less than 0.1 is suggested, as the criteria for a measured magnetic field being considered as being force-free. This criteria has been adopted and used by the community in all afterwards studies as well as in ours.

By studying the spectropolarimter data from Mees Solar Observatory using the Na I $\lambda$5896 spectral line, \citet{Metcalf1995} concluded that NOAA 7216 is not force-free in the photosphere but becomes force-free at a height of about 400 km and above in the chromosphere.  However, also using the Mees spectropolarimter data but on the Fe 6301.5/6302.5 lines,  \citet{Moon2002} analyzed 12 vector magnetograms of three flare-eruptive active regions and concluded that the photospheric magnetic fields are not far away from being force-free. The median value of $|F_z/F_p|$ that \citet{Moon2002} found is $0.13$, which is far smaller than the value (around $0.5$) that  \citet{Metcalf1995} obtained for the photospheric magnetic field.  \citet{Tiwari2012} studied a few high spatial resolution SP/Hinode magnetograms, and also concluded that the sunspot magnetic fields are not far away from being force-free, although their force-freeness may change with the time.

Disputable results continue to appear. \citet{Liu2013} carried out a statistical study by analyzing a sample of 925 magnetograms which were taken with the filter-based Solar Magnetic Field Telescope (SMFT) of the Huairou Solar Observing Station (HSOS). They found that only about $25\%$ of the magnetograms can be considered as close to be force-free, i.e., most ($75\%$) of the photospheric magnetic fields are not force-free. Yet when \citet{Duan2020} studied a sample of 3536 HMI/SDO magnetograms of 51 flux emerging active regions, they concluded that the emerging active regions are ``very close to a force-free state, which is not consistent with theories as well as idealized simulations of flux emergence''.

Different from above studies that directly applied Equations (2)-(5) to observed magnetograms and gave their respective conclusions, \citet{ZhangXM2017} tried to reconcile the differences in the results of previous studies by carrying out a systematic study on how the instrumental effects could influence the judgement of the force-freeness. They studied how the limited field of view, the instrument sensitivity and the measurement noise could affect the force-freeness judgement. They found that as long as the flux balance condition is roughly satisfied, the influences of the field of view and the instrument sensitivity on the force-freeness judgement are not large. However, the influence of the measurement noise on the force-freeness judgement is huge. It could make a truly force-free field be wrongly judged as being not-force-free and a truly not-force-free field be wrongly judged as being force-free. This is due to the fact that the noise level of the transverse field is about or even more than ten times of  that of the vertical field in the magnetograms \citep{Wiegelmann2012, Hoeksema2014}. The square of the noises of the transverse fields could not cancel out those of the vertical fields when using Equation (4), a point discussed in detail in the discussion section of \citet{ZhangXM2017}. They also pointed out that cutting the magnetograms at a high signal-to-noise level would help some but cannot overturn the result. Their examples show that even when cutting the magnetograms at a $2\sigma$ level (which is the case in most studies), the noises in the remaining $10\%$ data points can still affect the force-freeness judgement.

The study of \citet{ZhangXM2017} may explain why previous studies gave disputable results. It may merely depend on how the authors have handled the measurement noises. In other words, the measurement noises may have influenced the force-freeness judgments in a way that has been largely underestimated in most studies. In this paper, we propose an approach to overcome this problem: by reducing the spatial resolution to lower the noise level. In doing so, the heavy influence of the measurement noise on the force-freeness judgement will be significantly suppressed. We first use two analytical magnetic field solutions, in Section 2, to show the success and the effectiveness of this approach. Then we apply this approach to two large datasets of active region magnetograms. The data and the samples will be described in Section 3. The analysis and results will be presented in Section 4. A brief summary and discussion will be given in Section 5.

\section{The Model} 

In this section we use two analytical solutions, one known force-free magnetic field, described in \citet{Low1990}, and one known force-balanced non-force-free magnetic field, described in \citet{Low1992}, to show the effectiveness of our proposed approach: reducing the spatial resolution to lower the noise level so that the heavy influence of the measurement noise on the force-freeness judgement could be significantly suppressed. Similar approach has been proposed and tested in \citet{Jiang2018}. For the completion of this paper, we present it here using two different solutions from theirs.

The top two panels of Figure 1 show the two vector magnetic fields we used. The one in the left top panel is constructed using nonlinear force-free solutions described in \citet{Low1990}. It is the same to the one presented in the Figure 4 of  \citet{Low1990}, generated by $P_{1,1}$ with $l=0.3$ and $\Phi=\pi/4$. Red contours outline the positive vertical magnetic fields ($B_z>0$) and the blue lines outline the negative vertical magnetic fields ($B_z<0$). The green lines show where the $B_z=0$ lies. The arrows in the plane show the directions of the transverse fields ($\bf{B_t}$), with their lengths in proportion to the magnitudes of the transverse field ($|\bf{B_t}|$). 

The constructed magnetogram has a size of $1024\times1024$ pixels$^2$. Following \citet{ZhangXM2017}, we have rescaled the vector magnetogram to make its maximum strength of $B_z$ to be $2000 ~G$. This makes the field strengths in the obtained magnetogram more close to the observed ones. Specifically, a factor of 6.80983 has been multiplied to $B_x$, $B_y$ and $B_z$, making the minimum of $B_z$ changing from its original $-293.693 ~ G$ value in \citet{Low1990} solution to $-2000 ~G$. Using Equations (2)-(5), we get $F_x/F_p=-0.000039$, $F_y/F_p=0.000649$ and $F_z/F_p=-0.002797$, verifying its nature as being force-free. 

In order to use Equations (2)-(5) to estimate the force-freeness, the magnetogram needs to have its positive and negative fluxes roughly balanced.   The quantity, flux imbalance ($MI$), defined as 
\begin{equation}
MI = \frac{|F^+ - F^-|}{F^+ + F^-} \times 100 \% ~~,
\end{equation}
\noindent where $F^+ $and $F^-$ are upward ($B_z>0$) and downward ($B_z<0$) magnetic fluxes respectively, is usually estimated and used as the indicator of the degree of flux balance.  Most previous studies use a criteria of $10\%$, that is, magnetogram whose magnetic imbalance ($MI$) is within $10\%$ is regarded as satisfying the prerequisite for applying the \citet{Low1985} method. Our previous study \citep{ZhangXM2017} has also verified the applicability of this criteria. The $MI$ value of this \citet{Low1990} solution field is $-3.09\%$.

Presented in a similar way in the right top panel of Figure 1 is a vector magnetic field constructed using the magnetostatic solutions described in \citet{Low1992}. The vertical magnetic field ($B_z$) is obtained using model parameters listed in the Table 1 of \citet{Low1992}, making the $B_z$ similar to that presented in the Figure 1 of \citet{Low1992}, except that the field of view has been shifted a little bit up in the y-direction in order to make the obtained field possessing a value of $MI$ less than $10\%$. For other model parameters in \citet{Low1992}, we use $\alpha_0=0.3$ and $a=0.5$.  Similarly, the constructed vector magnetogram has a size of $1024\times1024$ pixels$^2$ and has been rescaled to make its maximum strength of $B_z$ to be $2000 ~G$. This field has $F_x/F_p=0.002630$, $F_y/F_p=-0.020666$ and $F_z/F_p=-0.144740$. So, by its $F_z/F_p$ value, it is a non-force-free field. Its $MI$ is $0.635\%$. 

The remaining six panels of Figure 1 show how reducing the spatial resolution could influence the force-freeness judgement. The left panels are of the \citet{Low1990} force-free field, and the right panels are of the \citet{Low1992} non-force-free field. The x-axis is the resolution-index $R$, where $R=0$ means that the original spatial resolution (magnetograms of $1024\times1024$ pixels$^2$) is used to calculate the $F_x/F_p$, $F_y/F_p$ and $F_z/F_p$. $R=1$ means that the original $1024\times1024$ pixels$^2$ magnetograms have been ``rebinned'' (using IDL program) to $512\times512$ pixels$^2$ magnetograms to calculate the $F_x/F_p$, $F_y/F_p$ and $F_z/F_p$. Similarly, $R=2, 3, 4$ means that the magnetograms have been further rebinned to $256\times256$ pixels$^2$, $128\times128$ pixels$^2$ and $64\times64$ pixels$^2$ magnetograms, respectively, to calculate the $F_x/F_p$, $F_y/F_p$ and $F_z/F_p$. In short, from $R=0$ to $R=4$, in each step, the spatial resolution of the magnetograms is reduced to $1/2$ of the previous ones, and the spatial resolution in $R=4$ magnetograms is only $1/16$ of the original ones. Similar practice is done for all reducing spatial resolution processes in the later development of the paper.

We can see from these six panels of Figure 1 that, even though the spatial resolutions of the magnetograms have been greatly reduced, the values of $F_x/F_p$, $F_y/F_p$ and $F_z/F_p$ do not change much. Take the bottom left panel for example, the $F_z/F_p$ values change from $-0.002797$ in the $R=0$ magnetogram to $-0.002594$ in the $R=4$ magnetogram, keeping correctly ``judging'' that the studied magnetic field is force-free. Similarly, in the bottom right panel,  $F_z/F_p$ values change from $-0.144740$ in the $R=0$ to $-0.144711$ in the $R=4$, keeping correctly ``judging'' that the studied magnetic field is not force-free. This is understandable, because the formula we use to calculate the $F_x$, $F_y$ and $F_z$ are all integrals on the whole field of view. Reducing the spatial resolution will certainly change the field magnitude in each data point, but the changes in the integrals may not be that large, particularly what we calculated are their ratios to $F_p$.

Note that above analysis are all done on the analytical solution fields which have no measurement noises. In reality, the used magnetograms, inverted from polarization measurements, all possess measurement noises. In particular, the noise level of the transverse field is about ten times of that of the vertical field in the magnetograms \citep{Wiegelmann2012, Hoeksema2014}. This will bring in serious problem as discussed in \citet{ZhangXM2017}. In Figure 2 we present this influence again and in particular we show how reducing the spatial resolution to lower the noise level could significantly reduce this bad influence of the measurement noise on the force-freeness judgement.
 
Similar to Figure 1, the left panels in Figure 2 present the $F_x/F_p$, $F_y/F_p$ and $F_z/F_p$ values of the \citet{Low1990} force-free field and the right panels are for the \citet{Low1992} non-force-free field. Following \citet{ZhangXM2017}, to mimic the effect of the measurement noise in the observed magnetogram, we have added a white noise to each component of the vector magnetogram. That is, the $B_x$, $B_y$ and $B_z$ component in the above two $1024\times1024$ pixels$^2$ magnetograms are replaced by $B_x+\sigma_x$, $B_y+\sigma_y$ and $B_z+\sigma_z$, respectively, where $\sigma_x$, $\sigma_y$ and $\sigma_z$ are the white noise of each field component. $\sigma_z$ in each data point is created by  multiplying $\sigma_z^0$ with a normally-distributed random number, similarly $\sigma_x$ and $\sigma_y$ by multifying $\sigma_x^0$ and $\sigma_y^0$ with a normally-distributed random number of their own. Also following \citet{ZhangXM2017}, to imitate the real magnetograms, we have assumed $\sigma_x^0$ and $\sigma_y^0$ = 10 $\sigma_z^0$, with $\sigma_z^0$ increasing from $0 ~G$ to $24 ~G$ with a constant step of $1 ~G$. The values of $\sigma_z^0$ are plotted as the x-axis in Figure 2.

The black lines in Figure 2 show the results of $F_x/F_p$ (top panels),  $F_y/F_p$ (middle panels) and $F_z/F_p$ (bottom panels) using above $1024\times1024$ pixels$^2$ noise-bearing magnetograms. We can see that, similar to what found in \citet{ZhangXM2017}, the measurement noises give little influence to $F_x/F_p$ and $F_y/F_p$. The variations of $F_x/F_p$ and $F_y/F_p$, with different noise levels, are all very minor. However, the $F_z/F_p$ increases monotonously with the increase of the noise level. When $\sigma_z^0$ is larger than $9 ~G$ ($90 ~G$ for $\sigma_x^0$ and $\sigma_y^0$), the $F_z/F_p$ in the left bottom panel has increased to a value larger than $0.1$. This means that, at this noise level and above, this real force-free magnetic field could be wrongly judged as being not-force-free. Similarly, When $\sigma_z^0$ is larger than $14 ~G$ ($140 ~G$ for $\sigma_x^0$ and $\sigma_y^0$), the $F_z/F_p$ in the right bottom panel has increased to a value larger than $-0.1$. This means that, at this noise level and above, this real non-force-free magnetic field could be wrongly judged as being force-free.

This serious problem will be significantly alleviated if we reduce the spatial resolutions of the magnetograms. In Figure 2, the blue lines show the $F_x/F_p$, $F_y/F_p$ and $F_z/F_p$ values for the $R=1$ magnetograms, where the noise-bearing $1024\times1024$ pixels$^2$ magnetograms have been rebinned to $512\times512$ pixels$^2$ magnetograms. We can see from the bottom left panel of Figure 2, now $\sigma_z^0$ needs to be as large as $17 ~G$ to make a wrong judgement. And from the bottom right panel of Figure 2, we see that no wrong judgement will be made now, even though the $F_z/F_p$ values still monotonously increase with the noise level. If we further decrease the spatial resolutions, the influences of the noise level on the force-free judgement will become less and less. This is evident in the red, green and yellow lines, which show the $F_x/F_p$, $F_y/F_p$ and $F_z/F_p$ values for the $R=2$, $R=3$ and $R=4$ magnetograms, respectively. We can see that, when $R=4$ is applied, the influence of the measurement noise has become almost neglectable. The two yellow lines in the bottom panels of Figure 2 have become almost horizontal. This means that when $R=4$ is applied, the estimated $F_x/F_p$, $F_y/F_p$ and $F_z/F_p$ values from the noise-bearing magnetograms have become nearly the same to their true values, independent of the noise level. This proves that reducing the spatial resolution can significantly suppress the bad influence of the measurement noise. 

Before we apply this method to two large datasets in the following sections of this paper, two notes are in order here. First, the analytic models we used here do not have small-scale structures as those in real data. With real data, the smoothing would change the basic magnetic-field structure in a larger degree. However, this will not change any conclusions presented in this section. This can be seen from the two examples in \citet{Jiang2018}. The two model fields there, reconstructed from observed magnetograms, do have small-scale structures. And similar analysis as those done in this section gives the same conclusion. An English version of \citet{Jiang2018} can be found in \citet{Jiang2019}. Secondly, we need to point out that our noise model, which fixes the ratio between the noise of the transverse field and that of the vertical field as a number of 10, is probably over-simplified. The real noise levels in magnetograms may vary from one to another and may also vary between data points and dependent on the field strength even within one magnetogram. Quoting from \citet{Wiegelmann2012} that ``the random errors in the line-of-sight component are about 5 G, while the uncertainty in the transverse field is as much as 200 G in weak field regions and as little as 70 G where the field is strong'', our approach of fixing the ratio as a number of 10 is indeed simple and somewhat conservative. However, in our regard, more accurate and more complicated noise models will not change our results in any qualitative way, if not even strengthen our conclusion further.

\section{The Data and Samples}

We construct two large datasets, one from HMI/SDO observations and one from SP/Hinode observations.

The first dataset uses the vector magnetograms taken by Helioseismic and Magnetic Imager (HMI)/SDO \citep{Scherrer2012}. HMI/SDO observes the full solar disk at Fe\,{\small I} $\lambda$6173 with a $4096 \times 4096$ CCD detector to study the oscillations and the magnetic fields on the solar photosphere. The spatial resolution is 0.91$''$ with a 0.5$''$ pixel size. A Milne–Eddington based inversion code \citep{Borrero2011} is used to derive the vector magnetograms from the filtergrams taken at six wavelength positions. The azimuthal $180^{\circ}$ ambiguity was resolved by the ``minimum energy'' algorithm \citep{Metcalf2006}. 

We use the hmi.sharp\_cea\_720s series of the active-region vector magnetograms \citep{Bobra2014}.  In this series of the data, the disambiguated vector magnetograms are deprojected using Lambert cylindrical equal area projection method, presented as ($B_r$, $B_{\theta}$, $B_{\phi}$) in heliocentric spherical coordinates which corresponds to ($B_z$, -$B_y$, $B_x$) in heliographic coordinates. Each region in this series is given a HARP number. From May 2010 to April 2020, the HARP number increases from $HARPNUM=1$ to $HARPNUM=7412$. For each HARPNUM during this period, we download the one whose longitude is closest to the central meridian. We end up with 3838 vector magnetograms downloaded. Among these 3838 magnetograms we further select those having a NOAA number assigned in the fits header (by the ``noaa\_ar'' keyword in the header). This reduces the number of selected magnetograms to be 1335. Finally, we only use those magnetograms whose $MI$ is less than $10\%$. This results in a sample of 547 HMI/SDO vector magnetograms. In these 547 magnetograms, the largest one has a field of view of $1715''\times452''$, the smallest field of view is $108''\times84''$. The median value of the field of view is around 62200 arcsec$^2$, roughly equivalent to $600\times410$ pixels$^2$ with $0.5''$/pixel.

The second dataset uses the vector magnetograms obtained by the Spectro-polarimeter (SP) aboard Hinode \citep{Kosugi2007}. Using a $0.16''\times164''$ slit, SP/Hinode obtains line profiles of two magnetically sensitive Fe lines at 630.15 and 630.25 nm and nearby continuum. The resolution of these magnetograms is about 0.32$''$/pixel for the fast maps we used. The SP/Hinode data are calibrated \citep{Lites2013} and inverted at the Community Spectro-polarimetric Analysis Center (CSAC, http://www.csac.hao.ucar.edu/). The inversion is based on the assumption of the Milne-Eddington atmosphere model and a nonlinear least-square fitting technique where the analytical Stokes profiles are fitted to the observed profiles. The inversion gives 36 parameters including the three components of magnetic field and the filling factor. The 180$\arcdeg$ azimuth ambiguity is resolved by setting the directions of the transverse fields most closely to a current-free field.

We searched the CSAC webpage and downloaded 448 magnetograms for active regions observed by the SP/Hinode from January 2006 to October 2019. Each of these 448 magnetograms is of one active region and is the one that is closest to the central meridian among all those observations of this active region. Two samples are constructed from these 448 magnetograms, depending on whether the ``true field strength'' or the ``flux density'' magnetograms are used. The ``true field strength'' magnetograms use the inversion-derived field strength $B$, field inclination $\gamma$ and field azimuth $\phi$ to get $B_x=B \sin(\gamma) \cos(\phi)$, $B_y=B \sin(\gamma) \sin(\phi)$ and $B_z=B \cos(\gamma)$. In the ``flux density" magnetograms, $B_x=\sqrt{f} \cdot B \sin(\gamma) \cos(\phi)$,  $B_y=\sqrt{f} \cdot B \sin(\gamma) \sin(\phi)$ and $B_z=f \cdot B \cos(\gamma)$, where $f$ is the filling factor. Following the hmi.sharp\_cea\_720s series procedure, these vector magnetic fields are transformed to local heliographic coordinates. Then as before, only those magnetograms with $MI$ less than 10\% are used. This makes our SP/Hinode ``flux density" sample contain 144 magnetograms and the  SP/Hinode ``true field strength'' sample contain 138 magnetograms. The reason that these two samples are of a different size is that, when calculating the $MI$ value of each magnetogram, we have used their respective $B_x$, $B_y$ and $B_z$ in either the ``flux density'' magnetogram or the ``true field strength'' magnetogram.

\section{Analysis and Results}  

In this section we present our analysis and results on the three samples we constructed: the HMI/SDO sample of 547 active regions, the SP/Hinode ``flux density'' sample of 144 active regions and the SP/Hinode ``true field strength'' sample of 138 active regions.

For each magnetogram in each sample, we first calculate their $F_x/F_p$, $F_y/F_p$ and $F_z/F_p$ values with the original spatial resolution (the $R=0$ magnetograms). Then, we decrease the spatial resolution of each magnetogram to $1/2$ of its original one and calculate the $F_x/F_p$, $F_y/F_p$ and $F_z/F_p$ values again but using these $R=1$ magnetograms.  Same procedure is done for the $R=2$, $R=3$ and $R=4$ magnetograms. That is, for each magnetogram in the samples, we have calculated the $F_x/F_p$, $F_y/F_p$ and $F_z/F_p$ values five times, with the magnetogram rebinned to lower and lower spatial resolutions, from $R=0$ (original spatial resolution) to $R=4$ ($1/16$ of the original spatial resolution).  Note that we have not done any ``cutting'' of the data points, that is, we did not remove the data points with low signal-to-noise ratios as most other researches do. This is because the results in the previous section have shown that reducing spatial resolution is a very efficient way to suppress the influence of measurement noise. Also, as mentioned before, \citet{ZhangXM2017} has showed that ``cutting''  is not a very efficient method and cannot fully remove the bad influence of the measurement noise.  

Figure 3 shows how the $F_x/F_p$, $F_y/F_p$ and $F_z/F_p$ values change with the spatial resolution. Here NOAA 11739 is selected as an example because it is a common active region in all three samples. The top panels of Figure 3 show the magnetograms of this active region, left panel of HMI/SDO data, middle panel of the SP/Hinode ``flux density'' magnetogram and the right panel of the SP/Hinode ``true field strength'' magnetogram.  We can see that the field of view of HMI/SDO magnetogram is somewhat different from that of the SP/Hinode magnetograms. That is because we have not done anything to change the HMI/SDO HARP field of view, neither to SP/Hinode field of view. So in our three samples, although there are many common active regions, the sizes of their field of views are usually different, between the HMI/SDO magnetograms and the SP/Hinode ones. 

The $F_x/F_p$, $F_y/F_p$ and $F_z/F_p$ values of this active region are plotted as green filled-diamonds in the remaining panels of Figure 3, linked by a purple line in each panel. Again, left panels are of the HMI/SDO magnetogram, middle panels of the SP/Hinode ``flux density'' magnetogram and the right panels of the SP/Hinode ``true field strength'' magnetogram. As those in Figure 1, the x-axis is the resolution-index $R$. We can see that, with the increase of $R$ (decrease of spatial resolution), the $F_x/F_p$ and $F_y/F_p$ values do not change much, for all three magnetograms. This is consistent with what we found in previous section, as shown in Figure 2.  

However, from the bottom three panels of Figure 3 we can see that, the $F_z/F_p$ values decrease significantly with the increase of $R$. For the HMI/SDO magnetogram, the $F_z/F_p$ values change from $0.046$ at the $R=0$ to $-0.25$ at the $R=4$. This means that, we could wrongly judge this field as being force-free if we use the $R=0$ magnetogram, while it is actually not force-free from its  $F_z/F_p=-0.25$ value using the $R=4$ magnetogram. Similarly, for the SP/Hinode ``flux density'' magnetogram, the $F_z/F_p$ values change from $-0.34$ at the $R=0$ to $-0.38$ at the $R=4$. Its $F_z/F_p<-0.1$ property confirms that this field is not force-free. The relatively small variation of  $F_z/F_p$ values with the $R$ indicates that the noise level in the SP/Hinode magnetograms are not large, particularly compared to those in the HMI/SDO magnetograms. Another advantage of the SP/Hinode observations is that it can give us the true field strength of the magnetic field rather than just the flux density. From the right bottom panel of Figure 3, we can see that the $F_z/F_p$ values change from $-0.54$ at the $R=0$ to $-0.61$ at the $R=4$, which means that the magnetic field of this active region is actually far away from being force-free.

Figure 4 shows the histograms of the $F_x/F_p$ (left panels), $F_y/F_p$ (middle panels) and $F_z/F_p$ (right panels) values of the 547 HMI/SDO magnetograms. Shown from top to bottom panels are the histograms of these values obtained from the $R=0$, $R=1$, $R=2$, $R=3$ and $R=4$ magnetograms, respectively. The blue histograms are of the active regions in the northern hemisphere and the green ones of the active regions in the southern hemisphere. The mean values of $F_x/F_p$ or $F_y/F_p$ or $F_z/F_p$ are calculated, shown by the vertical purple line in each panel with the number listed in the top corner of each panel. Also listed in each panel is the non-force-freeness number ($NF$), which shows the percentage of the active regions that have their $|F_x/F_p|>0.1$ in left panels or their $|F_y/F_p|>0.1$ in middle panels or their $|F_z/F_p|>0.1$ in right panels.

We can see that the mean values of $F_x/F_p$ and $F_y/F_p$ are all small, and do not change much with the increase of $R$. However, the mean values of $F_z/F_p$ decrease from $-0.021$ in the $R=0$ panel to $-0.210$ in the $R=4$ panel, a tendency same to that of the NOAA 11739 in Figure 3. Here we see again that the active regions could be judged as being force-free by their $|F_z/F_p|$ values in the $R=0$ magnetograms whereas they are actually not force-free when judging from the $R=4$ magnetograms. These mean values are also plotted in the left panels of Figure 3 as the blue filled-circles linked by red lines. 

From the right panels of Figure 3 we can also see that, although the mean values of $F_z/F_p$ decrease from $-0.021$ in the $R=0$ panel to $-0.210$ in the $R=4$ panel, the distributions of the histograms do not change much. Note that in previous studies \citep{Moon2002, Tiwari2012, Duan2020} where the authors claimed that their data are nearly force-free, some of their $|F_z/F_p|$ show values larger than 0.1. So in terms of the distribution of histograms, our results do not differ very much from most of the previous publications. It is the systematic shift of the whole distribution with the increase of $R$ shows the effect of reducing measurement noises.

From Figure 4, we can see that, even though both the mean values of $F_x/F_p$ and $F_y/F_p$ are small, they actually have very different distributions. Most active regions have their $|F_x/F_p|<0.1$. This can be seen from the relatively small values of $NF$ (from $7.9\%$ to $15.0\%$) listed in each of the left panels. However, for the $F_y/F_p$ values, even though their mean values are small (from $-0.007$ to $-0.010$), most active regions actually have their $|F_y/F_p|$ values larger than $0.1$. This can be seen from that their $NF$ numbers lie between $75.7\%$ to $78.4\%$. The $NF$ numbers of $F_z/F_p$ increase from $41.9\%$ in the $R=0$ panel to $75.1\%$ in the $R=4$ panel. When accounting for all three components of the net Lorentz force, only about $14\%$ of these HMI/SDO magnetograms have their $|F_x/F_p|<0.1$, $|F_y/F_p|<0.1$ and $|F_z/F_p|<0.1$.

From Figure 4 we can also see that, there is a north-south asymmetric in the $F_y/F_p$ distributions.  $F_y/F_p$ tends to be negative for active regions in the northern hemisphere and positive in the southern hemisphere. We will see this north-south asymmetry also exists in the $F_y/F_p$ values of the SP/Hinode ``flux density'' magnetograms (Figure 5), but becomes less evident in the SP/Hinode ``true field strength'' magnetograms (Figure 6).  The reasons for this north-south asymmetry as well as for the difference between using the  ``flux density'' magnetograms and the ``true field strength'' magnetograms are not known. This north-south asymmetry implies an existence of a net equator-ward force. A simple guess is that it is there to act against the meridional flow that drives the regions in the opposite hemispheres apart to the poles. To our knowledge, such an asymmetry has not been reported before. It deserves further studies to check and analyze.

Similar to Figure 4, Figure 5 shows the histograms of the $F_x/F_p$, $F_y/F_p$ and $F_z/F_p$ values, but for the 144 SP/Hinode ``flux density'' magnetograms. We see that the mean values of $F_x/F_p$ and $F_y/F_p$ are also very small, and also do not change much with the increase of $R$. The mean values of $F_z/F_p$ decrease from $-0.274$ in the $R=0$ panel to $-0.328$ in the $R=4$ panel, the same tendency as those in Figures 3 and 4. We also see that the $NF$ values of $F_x/F_p$ are relatively small (from $6.3\%$ to $8.3\%$), are large for $F_y/F_p$ (from $59.7\%$ to $63.2\%$), and become even larger for $F_z/F_p$ (from $90.3\%$ to $91.7\%$). When accounting for all three components of the net Lorentz force, none of these SP/Hinode ``flux density'' magnetograms have their $|F_x/F_p|<0.1$, $|F_y/F_p|<0.1$ and $|F_z/F_p|<0.1$.

Similarly, Figure 6 shows the histograms of the $F_x/F_p$, $F_y/F_p$ and $F_z/F_p$ values, but for the 138 SP/Hinode ``true field strength'' magnetograms. Again we see that the mean values of $F_x/F_p$ and $F_y/F_p$ are small, and also do not change much with the increase of $R$. The mean values of $F_z/F_p$ decrease from $-0.406$ in the $R=0$ panel to $-0.471$ in the $R=4$ panel, the same tendency as those in Figures 3, 4 and 5. We also see here that the $NF$ values of $F_x/F_p$ are relatively small (from $9.4\%$ to $11.6\%$), are large for $F_y/F_p$ (from $68.1\%$ to $71.0\%$), and become even larger for $F_z/F_p$ (from $97.1\%$ to $98.6\%$). Again, when accounting for all three components of net Lorentz force, none of these SP/Hinode ``true field strength'' magnetograms have their $|F_x/F_p|<0.1$, $|F_y/F_p|<0.1$ and $|F_z/F_p|<0.1$. Most notably here is that, the mean values of $F_z/F_p$ are all smaller than $-0.4$. The mean value of  $F_z/F_p$ for the $R=4$ magnetograms becomes as low as $-0.471$, a number that is very close to what \citet{Metcalf1995} found in their study.

\section{Summary and Discussion} 

In a previous study \citep{ZhangXM2017} it was pointed out that, the much larger noise level of the transverse field in the observed magnetograms could bring in serious problem in the force-freeness judgement. It could make a truly force-free field be wrongly judged as being non-force-free and could also make a truly non-force-free field be wrongly judged as being force-free. 

In this paper we have proposed an approach to overcome this serious influence of the measurement noise. We proposed to reduce the spatial resolution of the magnetograms to lower the noise level so that the serious influence of the measurement noise on the force-freeness judgement could be significantly suppressed.  Using two analytical magnetic fields, one being force-free and one being non-force-free, we have shown that our proposed approach is very effective. The bad influence of the measurement noise could be largely removed when reducing the spatial resolution of the magnetograms to 1/16 of the original ones. 

We then applied this approach to three samples, constructed from two datasets of the HMI/SDO active region magnetograms and the SP/Hinode active region magnetograms, respectively.  Our analysis has shown that all three samples give consistent results that the photospheric magnetic field is not force-free. Most, if not all, active regions have one of their $|F_x/F_p|$ or $|F_y/F_p|$ or $|F_z/F_p|$ greater than $0.1$. In particularly, the mean value of $F_z/F_p$ is as low as $-0.47$ when using the SP/Hinode ``true field strength'' magnetograms, a value very close to what firstly found in \citet{Metcalf1995} for the photospheric magnetic field.

It is worthy of mentioning here that the non-force-freeness of the photospheric magnetic field does not exclude the applicability of using observed photospheric magnetograms as the boundary conditions for nonlinear force-free field extrapolations. Not only because the photospheric vector magnetograms are only those currently we can get with fair accuracy and reasonable spatial and temporal resolutions, but also because up-to-date extrapolation methods have taken into account this non-force-freeness fact by modifying the observed magnetograms to be force-free \citep{Wiegelmann2006,  Wiegelmann2010,  Wiegelmann2012}. These modified magnetograms should be regarded as presenting the chromospheric magnetic fields rather than the photospheric ones. 

At the same time,  ``many nonlinear force-free coronal-field extrapolations based directly on the photospheric vector magnetograms generally agrees with each other to some extent'' \citep{Duan2020} cannot be used as a proof that the photospheric magnetic field is close to be force-free. However, using the observed force-freeness to test and constrain flux emergence models and simulations is indeed a very interesting practice. Our preliminary study shows that, except for the very early stage of the flux emergence, the emerged active region quickly settles to $F_z/F_p<-0.1$, rather than the large positive  $F_z/F_p$ values shown in \citet{Toriumi2017} and \citet{Toriumi2020}. The point that an inconsistence exists between the observation and the simulation has already been touched by \citet{Duan2020}. It certainly worths a future detailed study on the evolution of the force-freeness in emerging active regions, using the method carried out in this paper, to test the consistency between the observation and various flux emergence simulations.

\begin{acknowledgements}
We thank the anonymous referee for helpful comments and suggestions that improved the presentation of this paper. The HMI data used in this paper were provided by courtesy of NASA/SDO and the HMI science team. Hinode is a Japanese mission developed and launched by ISAS/JAXA, collaborating with NAOJ as a domestic partner, NASA and STFC (UK) as international partners. The Hinode SOT/SP data used in this paper were distributed by the Community Spectropolarimetric Analysis Center of HAO/NCAR.  This work is supported by the National Natural Science Foundation of China (grant No. 11973056) and the National Key R\&D Program of China (grant No. 2021YFA1600500).
\end{acknowledgements}

\bibliography{ms}{}

\begin{thebibliography}{}
\expandafter\ifx\csname natexlab\endcsname\relax\def\natexlab#1{#1}\fi
\providecommand{\url}[1]{\href{#1}{#1}}
\providecommand{\dodoi}[1]{doi:~\href{http://doi.org/#1}{\nolinkurl{#1}}}
\providecommand{\doeprint}[1]{\href{http://ascl.net/#1}{\nolinkurl{http://ascl.net/#1}}}
\providecommand{\doarXiv}[1]{\href{https://arxiv.org/abs/#1}{\nolinkurl{https://arxiv.org/abs/#1}}}

\bibitem[{{Bobra} {et~al.}(2014){Bobra}, {Sun}, {Hoeksema}, {Turmon}, {Liu},
  {Hayashi}, {Barnes}, \& {Leka}}]{Bobra2014}
{Bobra}, M.~G., {Sun}, X., {Hoeksema}, J.~T., {et~al.} 2014, \solphys, 289,
  3549, \dodoi{10.1007/s11207-014-0529-3}

\bibitem[{{Borrero} {et~al.}(2011){Borrero}, {Tomczyk}, {Kubo},
  {Socas-Navarro}, {Schou}, {Couvidat}, \& {Bogart}}]{Borrero2011}
{Borrero}, J.~M., {Tomczyk}, S., {Kubo}, M., {et~al.} 2011, \solphys, 273, 267,
  \dodoi{10.1007/s11207-010-9515-6}

\bibitem[{{Charbonneau}(2014)}]{Charbonneau2014}
{Charbonneau}, P. 2014, \araa, 52, 251,
  \dodoi{10.1146/annurev-astro-081913-040012}

\bibitem[{{Duan} {et~al.}(2020){Duan}, {Jiang}, {Toriumi}, \&
  {Syntelis}}]{Duan2020}
{Duan}, A., {Jiang}, C., {Toriumi}, S., \& {Syntelis}, P. 2020, \apjl, 896, L9,
  \dodoi{10.3847/2041-8213/ab961e}

\bibitem[{{Gary}(2001)}]{Gary2001}
{Gary}, G.~A. 2001, \solphys, 203, 71, \dodoi{10.1023/A:1012722021820}

\bibitem[{{Hoeksema} {et~al.}(2014){Hoeksema}, {Liu}, {Hayashi}, {Sun},
  {Schou}, {Couvidat}, {Norton}, {Bobra}, {Centeno}, {Leka}, {Barnes}, \&
  {Turmon}}]{Hoeksema2014}
{Hoeksema}, J.~T., {Liu}, Y., {Hayashi}, K., {et~al.} 2014, \solphys, 289,
  3483, \dodoi{10.1007/s11207-014-0516-8}

\bibitem[{{Jiang} \& {Mei}(2019)}]{Jiang2019}
{Jiang}, C.-q., \& {Mei}, Z. 2019, \caa, 43, 252,
  \dodoi{10.1016/j.chinastron.2019.04.009}

\bibitem[{{Jiang} \& {Zhang}(2018)}]{Jiang2018}
{Jiang}, C.~Q., \& {Zhang}, M. 2018, Acta Astronomica Sinica, 59, 39

\bibitem[{{Kosugi} {et~al.}(2007){Kosugi}, {Matsuzaki}, {Sakao}, {Shimizu},
  {Sone}, {Tachikawa}, {Hashimoto}, {Minesugi}, {Ohnishi}, {Yamada}, {Tsuneta},
  {Hara}, {Ichimoto}, {Suematsu}, {Shimojo}, {Watanabe}, {Shimada}, {Davis},
  {Hill}, {Owens}, {Title}, {Culhane}, {Harra}, {Doschek}, \&
  {Golub}}]{Kosugi2007}
{Kosugi}, T., {Matsuzaki}, K., {Sakao}, T., {et~al.} 2007, \solphys, 243, 3,
  \dodoi{10.1007/s11207-007-9014-6}

\bibitem[{{Lites} \& {Ichimoto}(2013)}]{Lites2013}
{Lites}, B.~W., \& {Ichimoto}, K. 2013, \solphys, 283, 601,
  \dodoi{10.1007/s11207-012-0205-4}

\bibitem[{{Liu} {et~al.}(2013){Liu}, {Su}, {Zhang}, {Deng}, {Gao}, {Yang}, \&
  {Mao}}]{Liu2013}
{Liu}, S., {Su}, J.~T., {Zhang}, H.~Q., {et~al.} 2013, \pasa, 30, e005,
  \dodoi{10.1017/pasa.2012.005}

\bibitem[{{Low}(1985)}]{Low1985}
{Low}, B.~C. 1985, in Measurements of Solar Vector Magnetic Fields, ed. M.~J.
  {Hagyard}, 49--65

\bibitem[{{Low}(1992)}]{Low1992}
{Low}, B.~C. 1992, \apj, 399, 300, \dodoi{10.1086/171925}

\bibitem[{{Low} \& {Lou}(1990)}]{Low1990}
{Low}, B.~C., \& {Lou}, Y.~Q. 1990, \apj, 352, 343, \dodoi{10.1086/168541}

\bibitem[{{Metcalf} {et~al.}(1995){Metcalf}, {Jiao}, {McClymont}, {Canfield},
  \& {Uitenbroek}}]{Metcalf1995}
{Metcalf}, T.~R., {Jiao}, L., {McClymont}, A.~N., {Canfield}, R.~C., \&
  {Uitenbroek}, H. 1995, \apj, 439, 474, \dodoi{10.1086/175188}

\bibitem[{{Metcalf} {et~al.}(2006){Metcalf}, {Leka}, {Barnes}, {Lites},
  {Georgoulis}, {Pevtsov}, {Balasubramaniam}, {Gary}, {Jing}, {Li}, {Liu},
  {Wang}, {Abramenko}, {Yurchyshyn}, \& {Moon}}]{Metcalf2006}
{Metcalf}, T.~R., {Leka}, K.~D., {Barnes}, G., {et~al.} 2006, \solphys, 237,
  267, \dodoi{10.1007/s11207-006-0170-x}

\bibitem[{{Moon} {et~al.}(2002){Moon}, {Choe}, {Yun}, {Park}, \&
  {Mickey}}]{Moon2002}
{Moon}, Y.~J., {Choe}, G.~S., {Yun}, H.~S., {Park}, Y.~D., \& {Mickey}, D.~L.
  2002, \apj, 568, 422, \dodoi{10.1086/338891}

\bibitem[{{Scherrer} {et~al.}(2012){Scherrer}, {Schou}, {Bush}, {Kosovichev},
  {Bogart}, {Hoeksema}, {Liu}, {Duvall}, {Zhao}, {Title}, {Schrijver},
  {Tarbell}, \& {Tomczyk}}]{Scherrer2012}
{Scherrer}, P.~H., {Schou}, J., {Bush}, R.~I., {et~al.} 2012, \solphys, 275,
  207, \dodoi{10.1007/s11207-011-9834-2}

\bibitem[{{Tiwari}(2012)}]{Tiwari2012}
{Tiwari}, S.~K. 2012, \apj, 744, 65, \dodoi{10.1088/0004-637X/744/1/65}

\bibitem[{{Toriumi} \& {Takasao}(2017)}]{Toriumi2017}
{Toriumi}, S., \& {Takasao}, S. 2017, \apj, 850, 39,
  \dodoi{10.3847/1538-4357/aa95c2}

\bibitem[{{Toriumi} {et~al.}(2020){Toriumi}, {Takasao}, {Cheung}, {Jiang},
  {Guo}, {Hayashi}, \& {Inoue}}]{Toriumi2020}
{Toriumi}, S., {Takasao}, S., {Cheung}, M. C.~M., {et~al.} 2020, \apj, 890,
  103, \dodoi{10.3847/1538-4357/ab6b1f}

\bibitem[{{Wiegelmann} \& {Inhester}(2010)}]{Wiegelmann2010}
{Wiegelmann}, T., \& {Inhester}, B. 2010, \aap, 516, A107,
  \dodoi{10.1051/0004-6361/201014391}

\bibitem[{{Wiegelmann} {et~al.}(2006){Wiegelmann}, {Inhester}, \&
  {Sakurai}}]{Wiegelmann2006}
{Wiegelmann}, T., {Inhester}, B., \& {Sakurai}, T. 2006, \solphys, 233, 215,
  \dodoi{10.1007/s11207-006-2092-z}

\bibitem[{{Wiegelmann} \& {Sakurai}(2021)}]{Wiegelmann2021}
{Wiegelmann}, T., \& {Sakurai}, T. 2021, Living Reviews in Solar Physics, 18,
  1, \dodoi{10.1007/s41116-020-00027-4}

\bibitem[{{Wiegelmann} {et~al.}(2012){Wiegelmann}, {Thalmann}, {Inhester},
  {Tadesse}, {Sun}, \& {Hoeksema}}]{Wiegelmann2012}
{Wiegelmann}, T., {Thalmann}, J.~K., {Inhester}, B., {et~al.} 2012, \solphys,
  281, 37, \dodoi{10.1007/s11207-012-9966-z}

\bibitem[{{Zhang} \& {Low}(2005)}]{Zhang2005}
{Zhang}, M., \& {Low}, B.~C. 2005, \araa, 43, 103,
  \dodoi{10.1146/annurev.astro.43.072103.150602}

\bibitem[{{Zhang} {et~al.}(2017){Zhang}, {Zhang}, \& {Su}}]{ZhangXM2017}
{Zhang}, X.~M., {Zhang}, M., \& {Su}, J.~T. 2017, \apj, 834, 80,
  \dodoi{10.3847/1538-4357/834/1/80}

\end{thebibliography}
\bibliographystyle{aasjournal}

\begin{figure}
\plotone{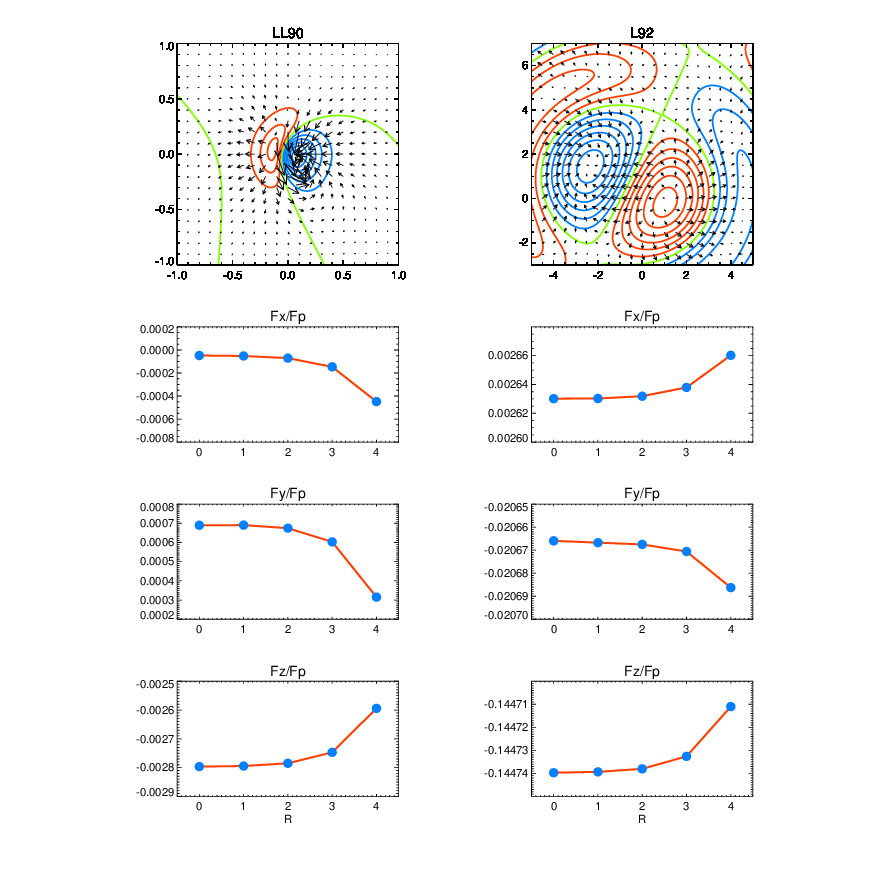}
\caption{From top to bottom panels present the magnetogram, $F_x/F_p$, $F_y/F_p$ and $F_z/F_p$ values for the analytical force-free magnetic field (left panels) and non-force-free magnetic field (right panels). The x-axis in the bottom six panels is the resolution-index $R$, with $R=0$ of the original spatial resolution. See text for more details.}
\end{figure}

\begin{figure}
\plotone{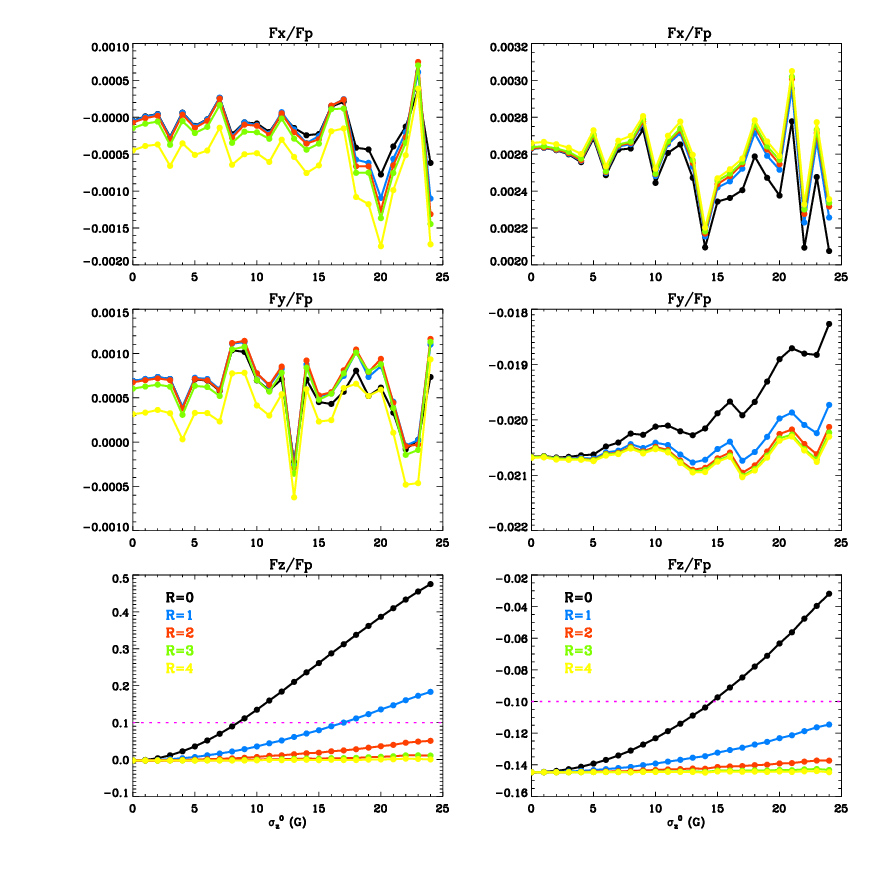}
\caption{From top to bottom panels present the $F_x/F_p$, $F_y/F_p$ and $F_z/F_p$ values for the analytical force-free magnetic field (left panels) and non-force-free magnetic field (right panels). The x-axis is the noise level in the vertical field ($\sigma_z^0$). Values obtained from magnetograms with different spatial resolutions are presented in different colors, black for R=0, blue for R=1, red for R=2, green for R=3 and yellow for R=4. See text for more details.}
\end{figure}

\begin{figure}
\plotone{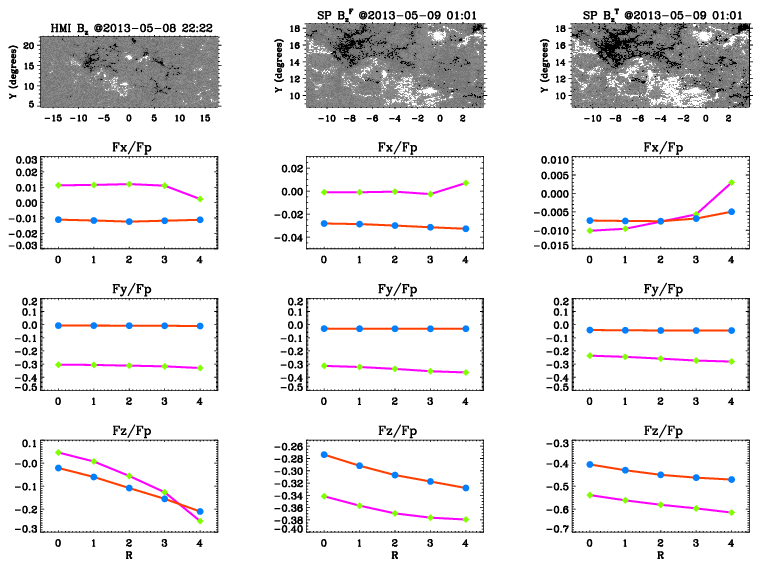}
\caption{From top to bottom panels present the magnetogram of NOAA 11739, $F_x/F_p$, $F_y/F_p$ and $F_z/F_p$ values. Left panels are of the HMI/SDO data, middle panels are of the SP/Hinode ``flux density'' data and right panels are of the SP/Hinode ``true field strength'' data. The x-axis in the bottom nine panels is the resolution-index $R$. The green filled-diamond points are of the NOAA 11739. The blue filled-circle points present the corresponding mean values in each sample (those also shown in Figures 4 - 6). See text for more details.}
\end{figure}

\begin{figure}
\plotone{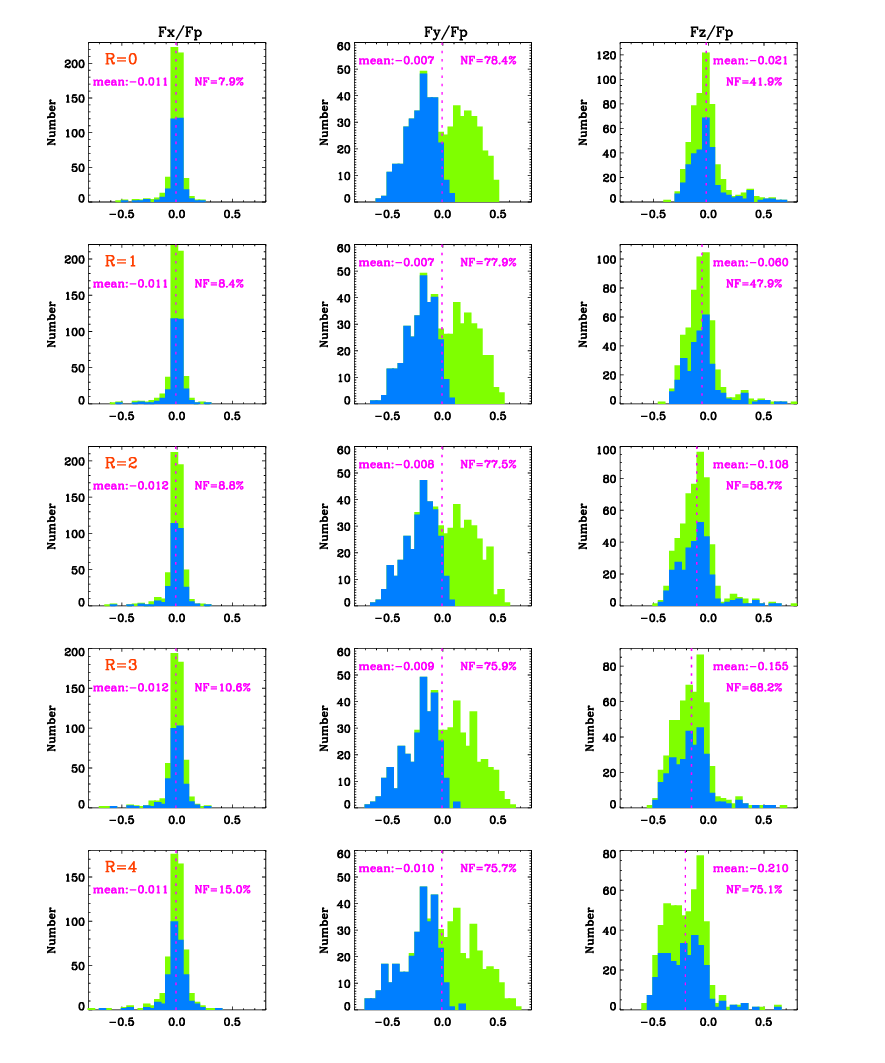}
\caption{The histograms of the $F_x/F_p$ (left panels), $F_y/F_p$ (middle panels) and $F_z/F_p$ (right panels) values of the 547 HMI/SDO magnetograms. From top to bottom panels are of the $R=0, 1, 2, 3, 4$ magnetograms respectively. Blue shows the portion of active regions in the northern hemisphere and green the portion of active regions in the southern hemisphere. Mean value and the $NF$ (Non-force-free percentage) value are also listed in each panel. Vertical purple line in each panel shows where the mean value lies.}
\end{figure}

\begin{figure}
\plotone{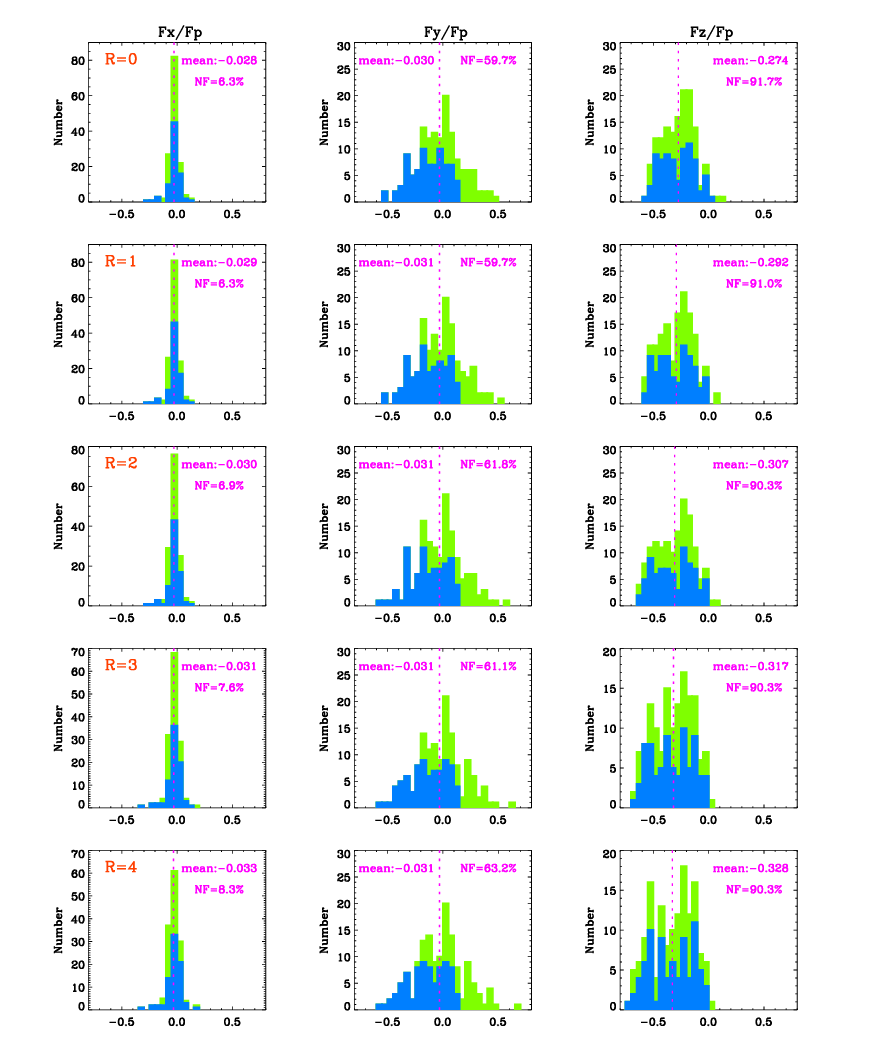}
\caption{Same as Figure 4 but for the sample of the 144 SP/Hinode ``flux density'' magnetograms.}
\end{figure}

\begin{figure}
\plotone{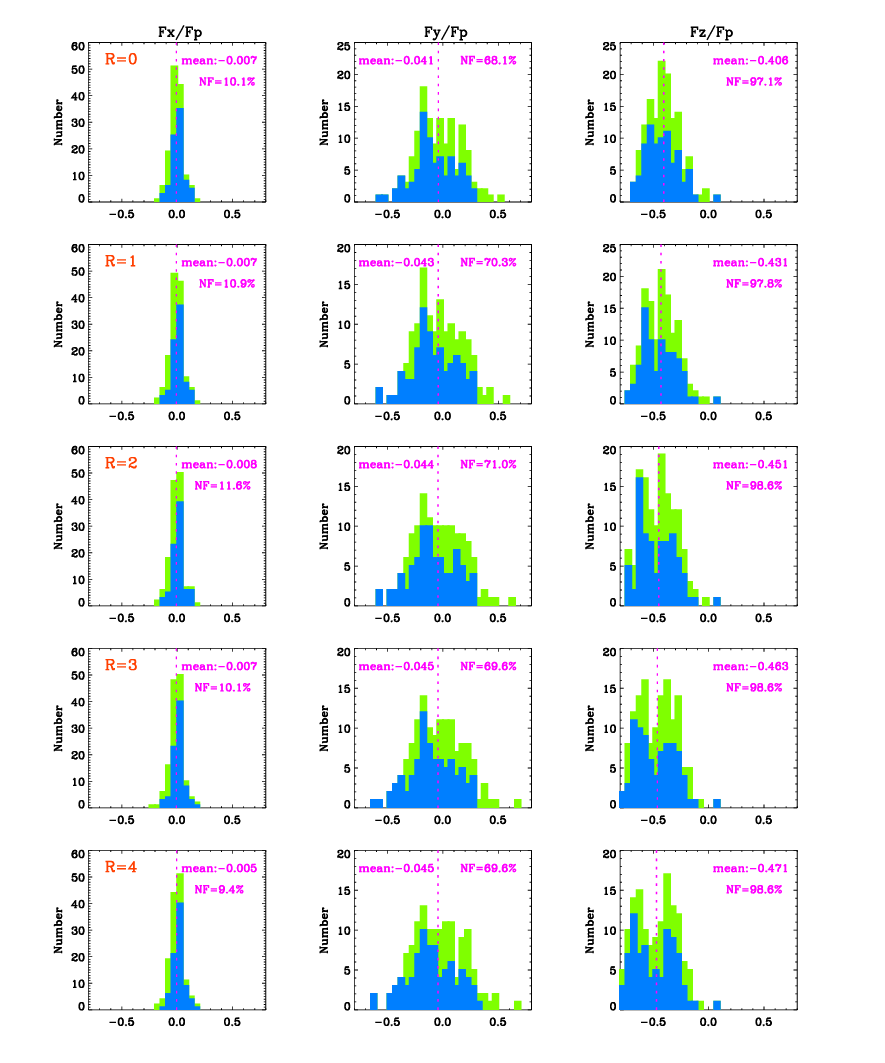}
\caption{Same as Figure 4 but for the sample of the 138 SP/Hinode ``true field strength'' magnetograms.}
\end{figure}

\end{document}